\documentclass[useAMS,usenatbib,longnamesfirst]{mn2e}

\usepackage{graphicx}
\bibpunct{(}{)}{;}{a}{}{,}

\newcommand\jcd{Christensen-Dalsgaard}
\newcommand\ea{et al.}
\newcommand{\hrd}{HR diagram}

\newcommand{\msun}{\ensuremath{M_\odot}}

\newcommand{\xc}{\ensuremath{X_\mathrm{c}}}
\newcommand{\dov}{\ensuremath{d_\mathrm{ov}}}

\newcommand{\teff}{\ensuremath{T_\mathrm{eff}}}

\newcommand{\Dzt}{\ensuremath{D_{02}}}
\newcommand{\DZT}{\ensuremath{\langle D_{02}\rangle}}
\newcommand{\Dot}{\ensuremath{D_{13}}}

\newcommand{\ETA}{\ensuremath{\eta}}
\newcommand{\THETA}{\ensuremath{\theta}}
\newcommand{\mc}{\ensuremath{m_\mathrm{core}}}
\newcommand{\scmc}{\ensuremath{(\Delta_\odot/\Delta)^2\mc/M}}

\def\myfigure#1#2#3#4{
        \begin{figure#4}
	\begin{center}
        \resizebox{\hsize}{!}{\includegraphics{#1}}
	\end{center}
        \caption{#2 \label{#3}}
        \end{figure#4}
}

\title[]{Asteroseismic Diagnostics of Stellar Convective Cores}
\author[A. Mazumdar et al.]{ Anwesh Mazumdar\thanks{E-mail:
anwesh.mazumdar@yale.edu}, Sarbani Basu, Braxton L.\ Collier and Pierre
Demarque\\
Astronomy Department, Yale University, P. O. Box 208101,
New Haven CT 06520-8101, U. S. A.}
\begin{document}

\date{Accepted . Received ; in original form }


\maketitle

\label{firstpage}

\begin{abstract}
We present a detailed study of the small frequency separations as
diagnostics of the mass of the convective core and evolutionary stage of
solar-type stars. We demonstrate how the small separations can be
combined to provide sensitive tests for the presence of convective
overshoot at the edge of the core. These studies are focused on low
degree oscillation modes, the only modes expected to be detected in
distant stars. Using simulated data with realistic errors, we find that
the mass of the convective core can be estimated to within 5\% if the
total stellar mass is known. Systematic errors arising due to
uncertainty in the mass could be up to 20\%. The evolutionary stage of
the star, determined in terms of the central hydrogen abundance using
our proposed technique, however, is much less sensitive to the mass
estimate.
\end{abstract}


\section{Introduction}
\label{sec:intro}

The theory of stellar structure predicts that the central region of
massive stars ($M\geq 1.1\msun$) are convective, rather than radiative
as is the case for the Sun.  Since convective flows imply chemical
mixing, the evolution of these stars is severely influenced by the
presence, and the extent, of the convective core.  The process of
convective transport also involves the question of overshoot which
implies the extension of the convective region beyond the classical
border of convective stability. This causes more hydrogen to be supplied
to the nuclear reactions going on in the core, thus effectively
increasing the lifetime of a star on the main sequence.  The present
theories of convection are inadequate to provide a definite measure of
the extent of overshoot and it is one of the free parameters in a
stellar model. The uncertainty in the mass of the convective core, or
its extension due to overshoot, directly translates into an error in the
determination of the age of the star. This in turn would affect other
studies based on stellar evolution like population synthesis in galaxies
etc. 

\myfigure{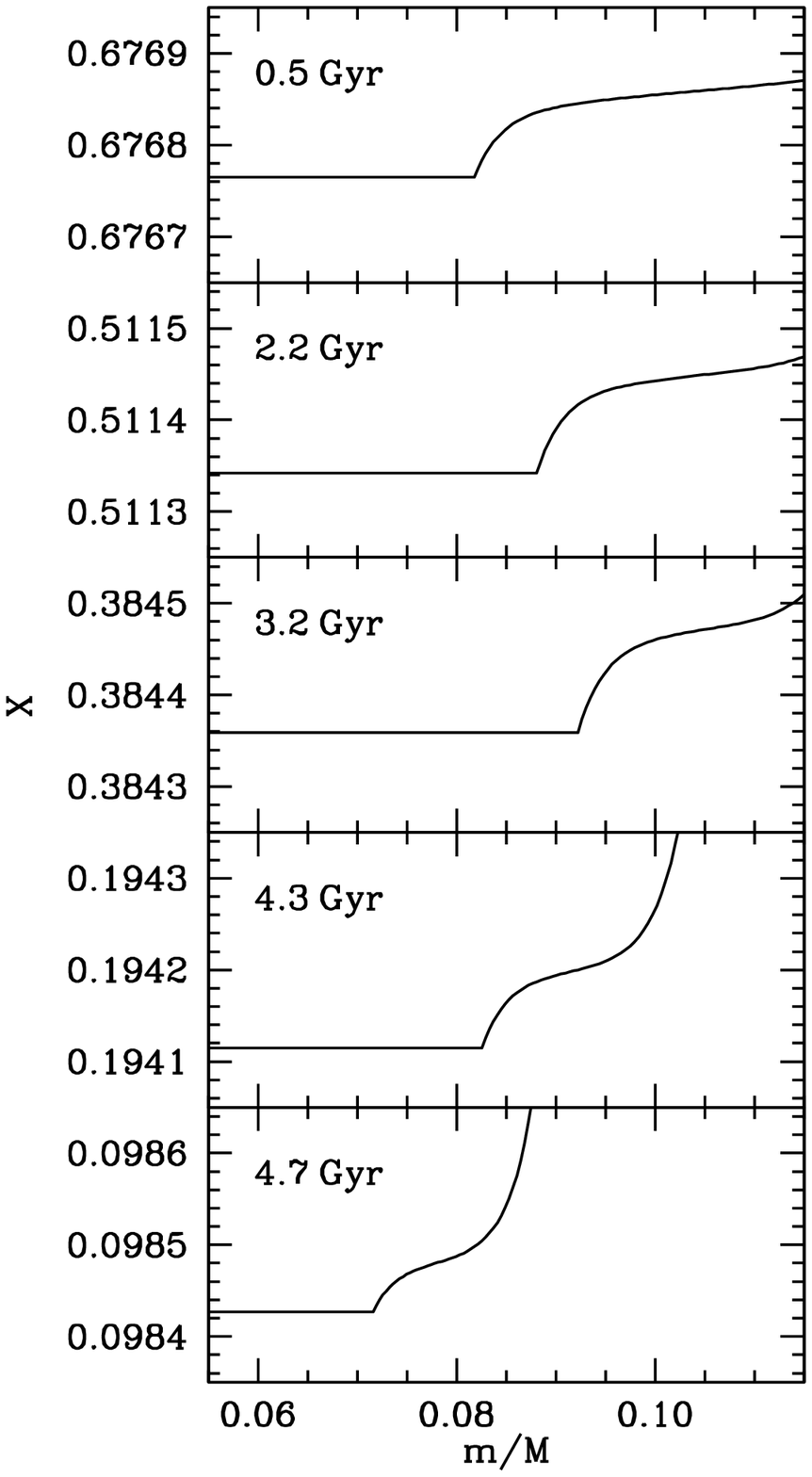}
{
The hydrogen profile near the boundary of the convective core is shown
for a $1.3\msun$ model with overshoot at different stages of evolution.
The flat portion of the curve indicates the core which is fully mixed
due to convection. It is clear that the size of the core increases
during the first part of evolution before shrinking in later stages.
Diffusion of hydrogen leaves a smoothed out step-like signature in the
hydrogen profile as the core recedes with increasing age of the star.
}
{fig:xprofile}
{}

The size of the convective core does not remain constant, and changes as
the star evolves.  The variation in the core size of a $1.3\msun$ star
is illustrated in Fig.~\ref{fig:xprofile} through the change in the
hydrogen profile as a function of fractional mass. For stars of mass
$1.1 \leq M/\msun \leq 2.0$ the core increases in size during the
initial stages of evolution before beginning to shrink later.
Furthermore, stellar models indicate that the extent and evolution of
the convective core is quite different depending on whether or not
convective overshoot is present (Fig.~\ref{fig:xc_mc}).  It is,
therefore, clear that a good test of the presence and extent of
convective layers in the central region of massive stars will go a long
way towards our understanding of stellar structure and evolution. While
there is no direct method to look inside a star, the recent advances in
asteroseismology does offer an opportunity to probe the innermost layers
of stars. The frequencies of oscillation of a star depend on its global
properties like the mass, radius etc, as well as the detailed internal
structure, especially the location of layers of rapid change in density
or chemical composition, and hence can be used to probe the inner
layers.

Space-based asteroseismic missions such as MOST \citep{wal03} and CoRoT
\citep{baglin03} are expected to measure the oscillation frequencies of
several stars with sufficient accuracy to enable a more detailed study
of stellar structure than has been possible so far. Ground-based
observations have already started measuring the frequencies of many
stars despite the difficulties of such measurements.  Unfortunately,
none of these observations can or will be able to observe the
intermediate degree modes  of oscillation that have been so useful in
solar studies. These missions can at most, hope to determine the
frequencies of modes with degree $\ell\le 3$. However, we are fortunate
in the fact that the modes of low degrees ($\ell=0$--$3$) penetrate the
central regions of a star and hence carry information about its deepest
layers.  We focus our study only on these modes.  

\myfigure{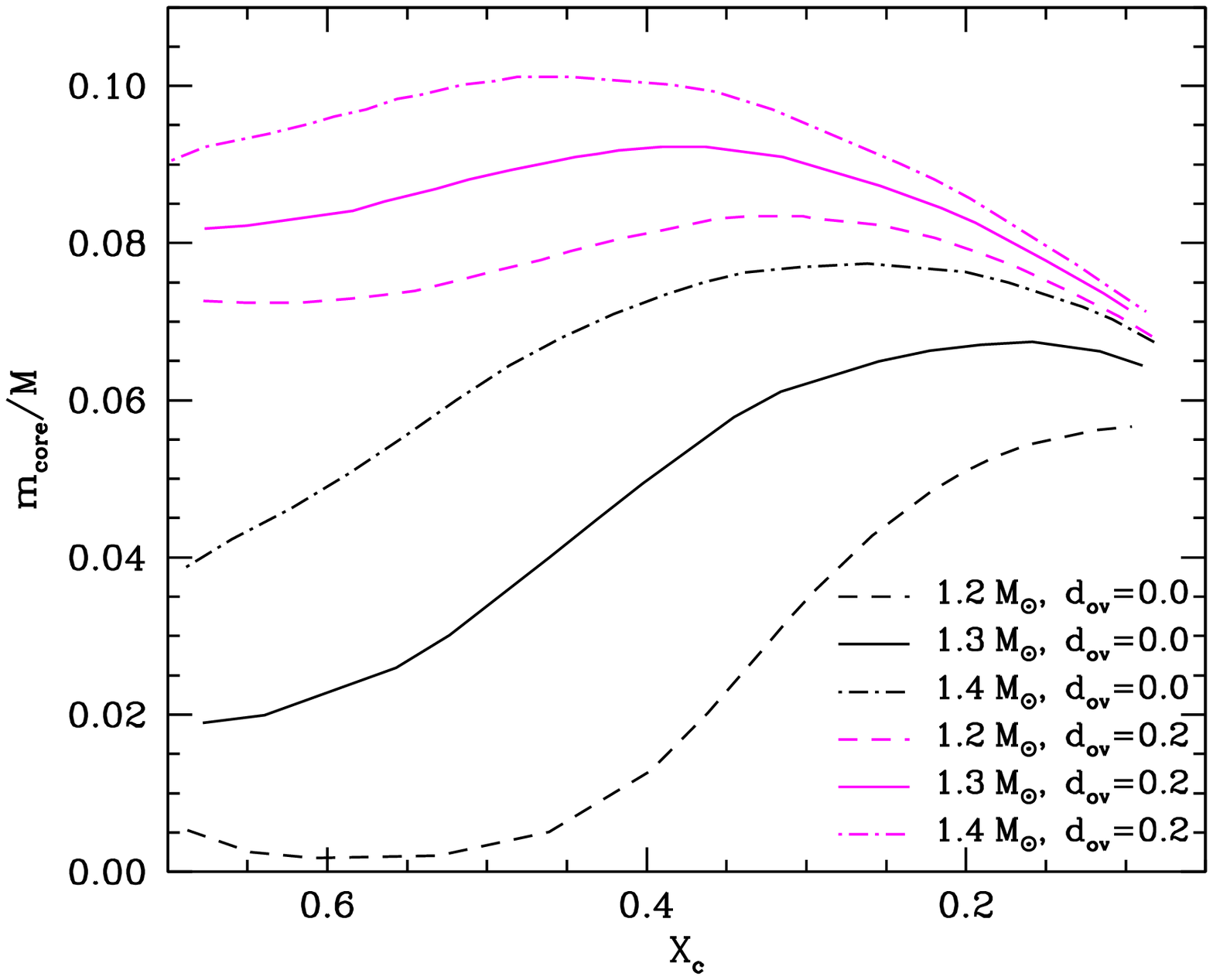}
{
The evolution of the fractional mass of the convective core, $\mc/M$ is
shown for models of mass $1.2$, $1.3$ and $1.4\msun$.  The black lines
represent models without convective core overshoot and the other curves
show models in which core overshoot ($0.2H_P$) has been incorporated. 
}
{fig:xc_mc}
{}

It has been suggested that the oscillatory signal in the frequencies due
to sharp features such as the boundaries of convective regions or
ionisation zones can be used to estimate the location of these layers
inside a star \citep{mct98,mct00}. While this technique appears to
succeed for the base of the outer convective envelope or the second
helium ionisation zone \citep{btg04,basu04}, \citet{ma01} have shown
that it would not work for the convective core due to a problem similar
to aliasing in Fourier analysis.  In this paper we show that
combinations of frequency separations can provide a better knowledge of
the core. 

It is well known that the large scale properties like the mass and
radius act mostly to scale the frequencies of nearly all modes of
oscillation. Such scaling effects are apparent in the large separation,
i.e., the difference in frequency of modes of successive order but of a
particular degree. The large separations are also quite sensitive to the
structure of the outer layers of a star, which are among the most
uncertain aspects of theoretical stellar models. On the other hand, the
so-called small separations, i.e., the small differences of frequency
between modes of nearly same order but different degrees, are more
affected by the deeper layers of a star. Therefore, the small
separations are useful in investigating the central features, especially
the extent of convective cores in massive stars.

Indeed, many efforts have been made to link the small frequency
separations to the convective core in massive stars \citep[for
theoretical studies see, e.g.,][]{dp91,rv01}. Using $\alpha$~Cen models,
\citet{gd00} had suggested to use the small separations to constrain the
age of the stars. After oscillations were indeed detected in
$\alpha$~Cen~AB \citep{bc02,cb03,bedding04,kjeldsen05}, different
authors have made use of the small separations to constrain the
parameters, especially the age, for these stars
\citep{thevenin02,thoul03,eggen04}.  Recently, \citet{rv03} have
demonstrated that the ratio of the small and large separations serve as
a better diagnostic of the innermost layers of a star than the small
separations alone.  \citet{mm05} have used this technique to refine the
seismic models for $\alpha$~Cen~AB. Similarly, the small separations
have proved to be useful for the seismic modelling of Procyon~A as well
\citep{ecb05,pbm06}.  \citet{straka05} have provided upper limits to the
convective overshoot in Procyon~A models using the frequency
separations.  In these studies, the frequency separations have been
mostly used to find the best possible model that fits the observed data
for a given star.  In this paper we aim to present a general technique
for probing specific properties of the central regions of a star, like
the mass of the convective core, and the state of evolution of the star,
using suitable combinations of the frequency separations.

While the large and small separation corresponding to each radial order
encodes information about a different layer inside the star, and is,
therefore, an independent diagnostic of the structure, there are also
advantages in considering suitable average values of these separations.
Firstly, the averaging over several radial orders removes the
small-scale variations due to subtle differences in the internal
structure, and enhances the effects of the more significant features of
the stellar interior. Second: it is often not possible from observations
to determine the frequency of modes of all radial orders for the same
degree, and an average is often the best indication of the trend of
frequencies.  Third: while individual frequencies, and hence the
separations are often susceptible to large observational errors, the
average values of the large and small separations turn out to be robust
quantities \citep{maz05}. The diagnostic power of the average
separations can be utilised very well through the \jcd\ diagram
\citep{jcd88} to determine the mass and age of a star. In this work, we
show that they can also be used for more detailed study of the stellar
interior and present a practical method involving the average values of
small and large separations and combinations thereof to estimate the age
of a star and the mass of the convective core.

In the next section we define new diagnostic quantities to probe the
stellar core and outline our technique and describe the stellar models
used in this work. Our results along with error estimates are presented
in Section~\ref{sec:results} with the concluding discussion in
Section~\ref{sec:discussion}.

\section[]{The Technique}
\label{sec:technique}

The average large separation of radial modes is defined as 
\begin{equation}
\Delta = \langle \nu(n+1,0) - \nu(n,0) \rangle,
\end{equation}
where the angular brackets imply averaging over a range of radial order,
$n$, appropriately chosen. 

The small separation between modes of degree $\ell$ and $\ell + 2$ is
defined as
\begin{equation}
\delta_{\ell\,\ell+2}(n) = \nu(n,\ell) - \nu(n-1,\ell+2)
\end{equation}
While comparing small separations between different pairs of modes
(e.g., $(\ell=0,2)$ and $(\ell=1,3)$) it is convenient to scale the
differences to eliminate the factor of $\ell(\ell+1)$ in the
frequencies. Thus one may define the scaled small separations as
\citep[cf.\ ][]{cb91,rv01}
\begin{equation}
D_{\ell\,\ell+2}(n) = \frac{\delta_{\ell\,\ell+2}(n)}{4\ell + 6} =
\frac{\nu(n,\ell) - \nu(n-1,\ell+2)}{4\ell + 6}
\end{equation}
In the rest of the paper, we will refer to this scaled definition,
$D_{\ell\,\ell+2}$, simply as the ``small separation''.  The small
separation between modes of degree $\ell=0$ and 2 are designated as
$D_{02}$, and those corresponding to $\ell=1$ and 3 as $D_{13}$.
\citet{rv00} have demonstrated that these two sets of small separations
encode similar, but somewhat different information about the central
layers. This is explained by the fact that modes of different degree
have different inner turning points, and therefore probe different
layers. Thus, it might be useful to compare the two small separations,
$D_{02}(n)$ and $D_{13}(n)$, of the same radial order. 

In order to exploit the fine differences between $D_{02}(n)$ and
$D_{13}(n)$, for each model, we define the following combinations of the
small separations:
\begin{equation}
\THETA (n_1,n_2) = \frac{\Delta_\odot}{\Delta}\,\left\langle D_{13}(n) -
D_{02}(n) \right\rangle_{n=n_1\,\mathrm{to}\,n_2}
\end{equation}
\begin{equation}
\ETA (n_1,n_2) =
\left(\frac{\Delta_\odot}{\Delta}\right)^2\,\left\langle (D_{13}(n) -
D_{02}(n))^2 \right\rangle_{n=n_1\,\mathrm{to}\,n_2}
\end{equation}
where the arguments $n_1$ and $n_2$ indicate the limits of radial orders
between which the averaging is carried out. $\Delta_\odot$ is the
average large separation of the Sun ($\Delta_\odot \simeq 135\mu$Hz).

The properties of the quantities \THETA\ and \ETA, and their sensitivity
to different stellar parameters, would partly depend on the range of the
radial order chosen for averaging. We deliberately choose relatively
higher order modes ($n \geq 10$) for averaging to avoid contamination
from mixed modes. In practice, however, the choice of the range for
averaging would be primarily governed by the availability of data in
different frequency ranges of the observed spectrum. For a complete set
of frequencies of a solar-type star, we find that the range of $n=15$ to
$28$ for \THETA\ and $n=10$ to $28$ for \ETA\ make these quantities
ideally sensitive to stellar properties such as the convective core
mass, \mc\ or the central hydrogen abundance, \xc. In the rest of the
paper, we shall assume these to be the respective ranges of averaging
for \THETA\ and \ETA, unless the range is stated explicitly.

We include the scaling factors $\Delta_\odot/\Delta$ in the definitions
of \THETA\ and \ETA\ to reduce homology effects while comparing stars of
different mass and radius. This factor essentially filters out the
global scaling of the frequencies due to differences in the mean density
of different stars, and enhances the effects of the structural details
which we want to study.

We study the variation of these new diagnostic quantities, \THETA\ and
\ETA\ with the age and evolution of the convective core in stars using a
large number of stellar models.  We restrict ourselves to stars with
masses just a little higher than the mass of the Sun, namely, the range
1.2--1.4\msun.  We study stellar models evolved to different ages on the
main sequence for masses $1.2$, $1.3$ and $1.4\msun$.  Ideally, the
boundary of the growing convective core should be tracked by adjusting
the mesh at each evolutionary step. However, since the current
evolutionary code does not have that capability, the edge of the core is
determined through the Schwarzschild criterion being tested over a very
fine mesh in the central regions of the star.  To compare with stars
without convection in their cores, we also constructed a sequence of
$1\msun$ models.  We also vary the convective overshoot in the core,
which is measured in units of the local pressure scale height, $H_P$. We
use two sets of models with the overshoot parameter $\dov =0$ and
$\dov=0.2$, where the amount of overshoot at the edge of the convective
core is given by $\dov H_P$.  For very small convective cores, the
overshoot is limited to be a fraction of the core size instead of $H_P$.
However, for most of the models that we consider, the size of the core
is typically at least an order of magnitude larger than the pressure
scale height at the boundary, so that we can simply use $\dov H_P$ as a
measure of the overshoot.

The models used in this work were constructed using YREC, the Yale
Rotating Evolution Code in its non-rotating configuration \citep{gue92}.
These models use the OPAL  equation of state \citep{rn02}, OPAL
opacities \citep{ir96}, low temperature opacities of \citet{af94} and
nuclear reaction rates as used by \citet{ba92}.  The models take into
account diffusion of helium and heavy elements, using the prescription
of \citet{tbl94}. The initial chemical composition of the models was
fixed at $(Y_0=0.26,Z_0=0.022)$.

\section[]{Results}
\label{sec:results}

\subsection{Testing for core overshoot}
\label{sec:res_ovshts}

\myfigure{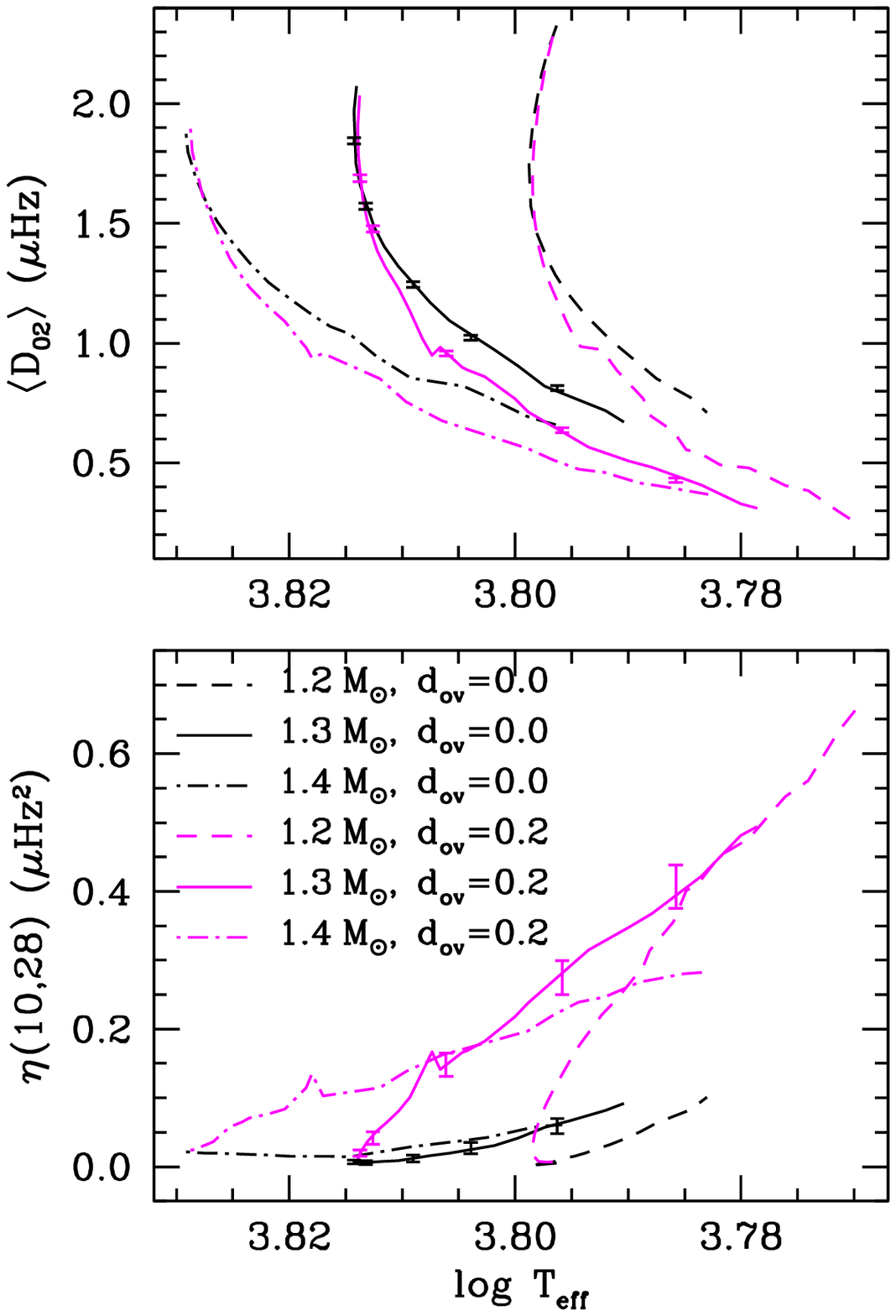}
{
The evolution of the average small separation \DZT\ ({\it upper panel})
and the new diagnostic \ETA\ ({\it lower panel}) as a function of the
effective temperature are shown. The black lines represent models
without overshoot while the other lines represent models with overshoot
($\dov = 0.2$). For each mass the curves for models with and without
overshoot diverge from the zero age main sequence.  The errorbars
represent $1\sigma$ errors in the respective quantities for relative
frequency errors of $10^{-4}$, obtained through simulations described in
Section~\ref{sec:res_err_random}. The significant kink in the curves,
especially for overshoot models, correspond to the age of the star when
the convective core stops expanding and begins shrinking. 
}
{fig:teff_d02_eta}
{}

As is clear from Fig.~\ref{fig:xc_mc} the evolution of the central
convective region of a star with age is significantly affected by the
presence of overshoot. Indeed, we find that this important difference is
reflected in the oscillation frequencies and consequently, the small
separations. The presence of convective overshoot, therefore, might be
tested through the small separations themselves. \citet{dimauro03} have
made a detailed investigation on how the frequency separations can be
used to distinguish between models with and without overshoot for a
post-main sequence star. \citet{maz05}, however, has shown that during
the main sequence phase the direct effect of overshoot on the small
separations $D_{02}$ becomes noticeable only in the later stages of
evolution.  We investigate whether the average small separations, \DZT,
or the quantity \ETA\ can provide any clue to the presence of overshoot
in a given star.   

In Fig.~\ref{fig:teff_d02_eta}, we show how  the average small
separations \DZT\ and the quantity \ETA\ vary as functions of the
effective temperature as the star evolves on the main sequence, for
models with and without core overshoot.  The errorbars represent the
$1\sigma$ errors in the ordinates for relative frequency errors of
$10^{-4}$. These errorbars were obtained from Monte Carlo simulations
with theoretical frequencies, described in detail in
Section~\ref{sec:res_err_random}.  It is evident that even if the mass
of a star is known, \DZT\ is barely sufficient to distinguish between
models with and without overshoot at a given \teff. The tracks for \ETA,
on the other hand, are well separated, with the overshoot models
producing much higher values of \ETA\ at a given \teff, irrespective of
mass. While the \DZT\ values for models with and without overshoot are
nearly the same, especially near the zero age main sequence (ZAMS), the
\ETA\ values of overshoot models diverge very rapidly beyond the ZAMS
from those of non-overshoot models. Thus, if the location of a star on
the \hrd\ is known, the quantity \ETA\ can be used to test the presence
of overshoot in the convective core. For large errors in  frequency the
errors in \ETA\ increase almost linearly (see Table~\ref{tab:err}); even
for relative errors $5\times 10^{-4}$ the separation between the tracks
is large enough to be distinctive to prove useful.

\subsection{Estimating the convective core mass, \mc}
\label{sec:res_mc}

\myfigure{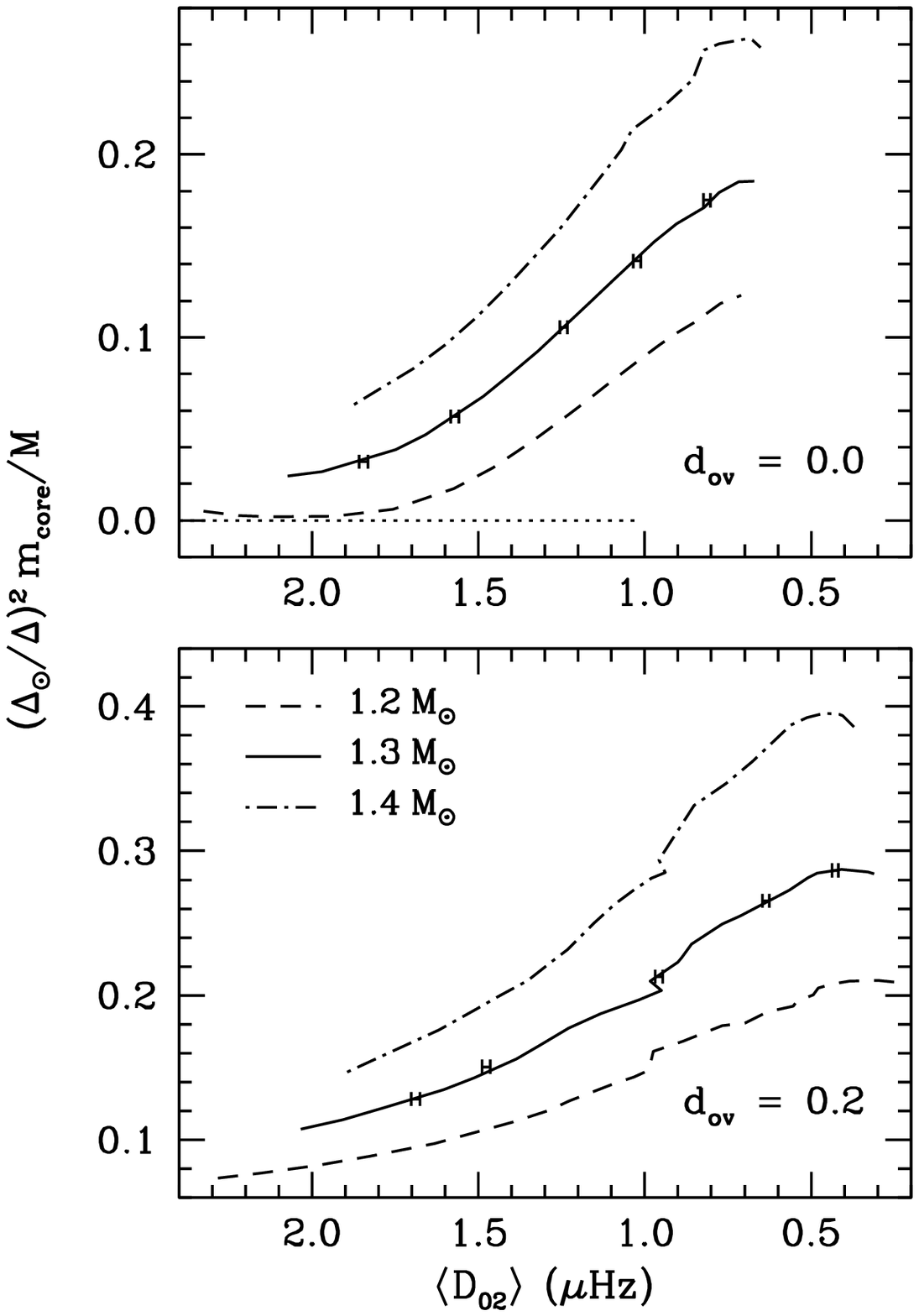}
{
The scaled mass of the convective core, \scmc\ is plotted as a function
of the average small separation, \DZT, for models without overshoot
({\it upper panel}) and with overshoot ({\it lower panel}). The dotted
line shows the \DZT\ values for a $1\msun$ model, for which the core is
not convective. The errorbars represent $1\sigma$ errors in \DZT\ for
relative frequency errors of $10^{-4}$, obtained through simulations
described in Section~\ref{sec:res_err_random}.  The significant kink in
the curves, especially for overshoot models, correspond to the age of
the star when the convective core stops expanding and begins shrinking.
}
{fig:d02_scmc}
{}

\myfigure{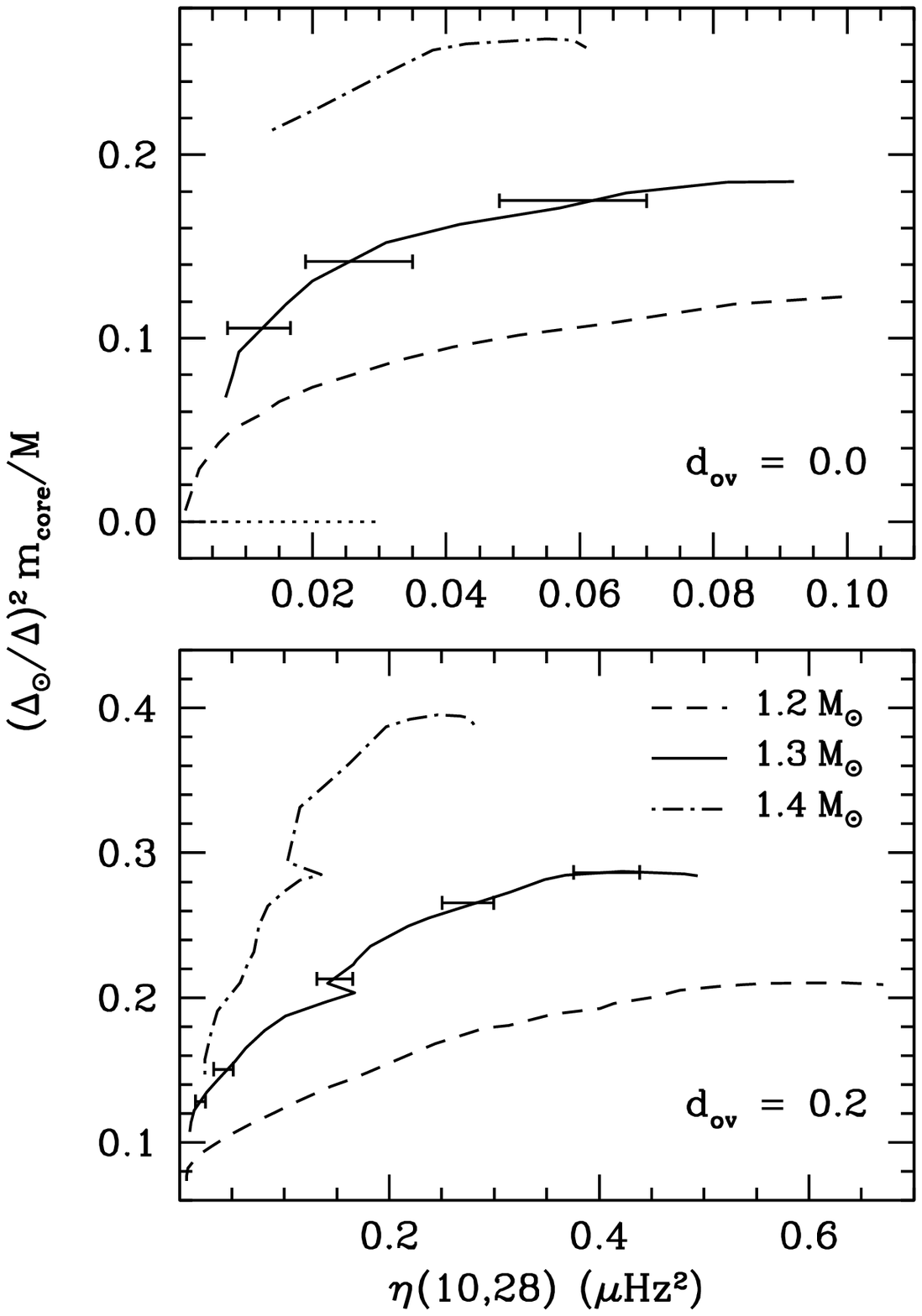}
{
The scaled mass of the convective core, \scmc\ is plotted as a function
of the diagnostic, \ETA, for models without overshoot ({\it upper
panel}) and with overshoot ({\it lower panel}). The dotted line shows
the \ETA\ values for a $1\msun$ model, for which the core is not
convective. The errorbars represent $1\sigma$ errors in \ETA\ for
relative frequency errors of $10^{-4}$, obtained through simulations
described in Section~\ref{sec:res_err_random}.  The significant kink in
the curves, especially for overshoot models, correspond to the age of
the star when the convective core stops expanding and begins shrinking.
}
{fig:eta_scmc}
{}

As evident from Figs.~\ref{fig:xprofile} and~\ref{fig:xc_mc}, the
fractional mass of the convective core of a star of total mass of about
$1.2$--$1.4$\msun\ does not remain constant throughout its main sequence
evolution. In particular, the convective core grows during the early
part of evolution and then begins to shrink at later stages. The exact
evolution of the core depends somewhat on the total mass and quite
strongly on whether or not convective overshoot occurs.  In either case,
the relationship between the fractional mass of the core and the age of
the star has opposite signs in two parts of evolution on the main
sequence.  Since the mass of the convective core is not monotonic with
the age, it is difficult to associate the core mass directly with the
small separations (and hence with \THETA\ or \ETA), which vary generally
monotonically with age.  However, we find that if we scale the mass of
the core with the factor $(\Delta_\odot/\Delta)^2$ which is related to
the mean density of the star, the scaled mass is nearly monotonic with
the age of the star. Of course, even this scaled mass of the core
eventually decreases with increasing age towards the end of the main
sequence when the core shrinks to a very small size. While theoretically
this might appear to be somewhat {\it ad hoc}, it provides a practical
way to overcome the difficult dichotomy of the core mass.
Observationally, it is a perfectly suitable method since the average
large separation can usually be measured to a fair degree of accuracy.
If the scaled core mass can be determined using a seismic diagnostic,
the actual mass of the core can easily be extracted through the average
large separation.

We have also found from Fig.~\ref{fig:teff_d02_eta} that the value of
\ETA\ is strongly dependent on the presence of overshoot. This implies
that we need separate calibration curves for models with and without
core overshoot. Since Fig.~\ref{fig:teff_d02_eta} itself provides a
means of testing for the presence of overshoot, from a practical
viewpoint, this does not pose a problem while dealing with a given star.
We use multiple models at different ages of $1.2$, $1.3$ and $1.4\msun$
both with and without overshoot to produce the needed calibration
curves.  

We first investigate the variation of the average small separation,
\DZT\ with the scaled fractional core mass, \scmc.  In
Fig.~\ref{fig:d02_scmc} we show the scaled core mass as function of
\DZT, for stellar models of different mass, evolving through the main
sequence. Every curve in this figure is an evolutionary track, which
serves as a calibration curve to determine \scmc\ for a given target.
The figure (and similarly Figs~\ref{fig:eta_scmc}, \ref{fig:eta_xc} and
\ref{fig:theta_xc}) have been drawn such that the sense of evolution is
from the left to the right along the $x$-axis. 

This figure illustrates the fact that although models with the same core
mass have largely different values of the small separation, the scaled
core mass is almost monotonic with \DZT. It is worth mentioning that the
kink in the curves shown in the lower panel of Fig.~\ref{fig:d02_scmc}
is not an artifact of the stellar models. It is, in fact, intrinsic to
the evolution of the star, and corresponds to the age when the increase
of the core size stops and the core begins to shrink (cf.\
Fig.~\ref{fig:xc_mc}). Such a kink is also seen in the \jcd\ diagrams
\citep{maz05}. The presence of the kink implies that a careful
calibration needs to be carried out -- perhaps two different
calibrations for two segments of the evolutionary track. For comparison,
we have also plotted the \DZT\ values for a $1\msun$ model, which does
not have a convective core at all. Every horizontal errorbar indicates
the $1\sigma$ error in \DZT\ due to relative frequency errors of
$10^{-4}$, which would translate into a random error in the
determination of \scmc\ through the local slope of the calibration
curves. Given the wide separation of the curves corresponding to
different masses, it is clear that such random error in estimating the
core mass due to errors in frequencies is much smaller than the
uncertainty due to the stellar mass. Therefore, to use this technique,
one needs to have a fair idea of the mass of the star. We shall
investigate both the random and systematic errors involved in detail in
Sections~\ref{sec:res_err_random} and~\ref{sec:res_err_sys}
respectively.

\myfigure{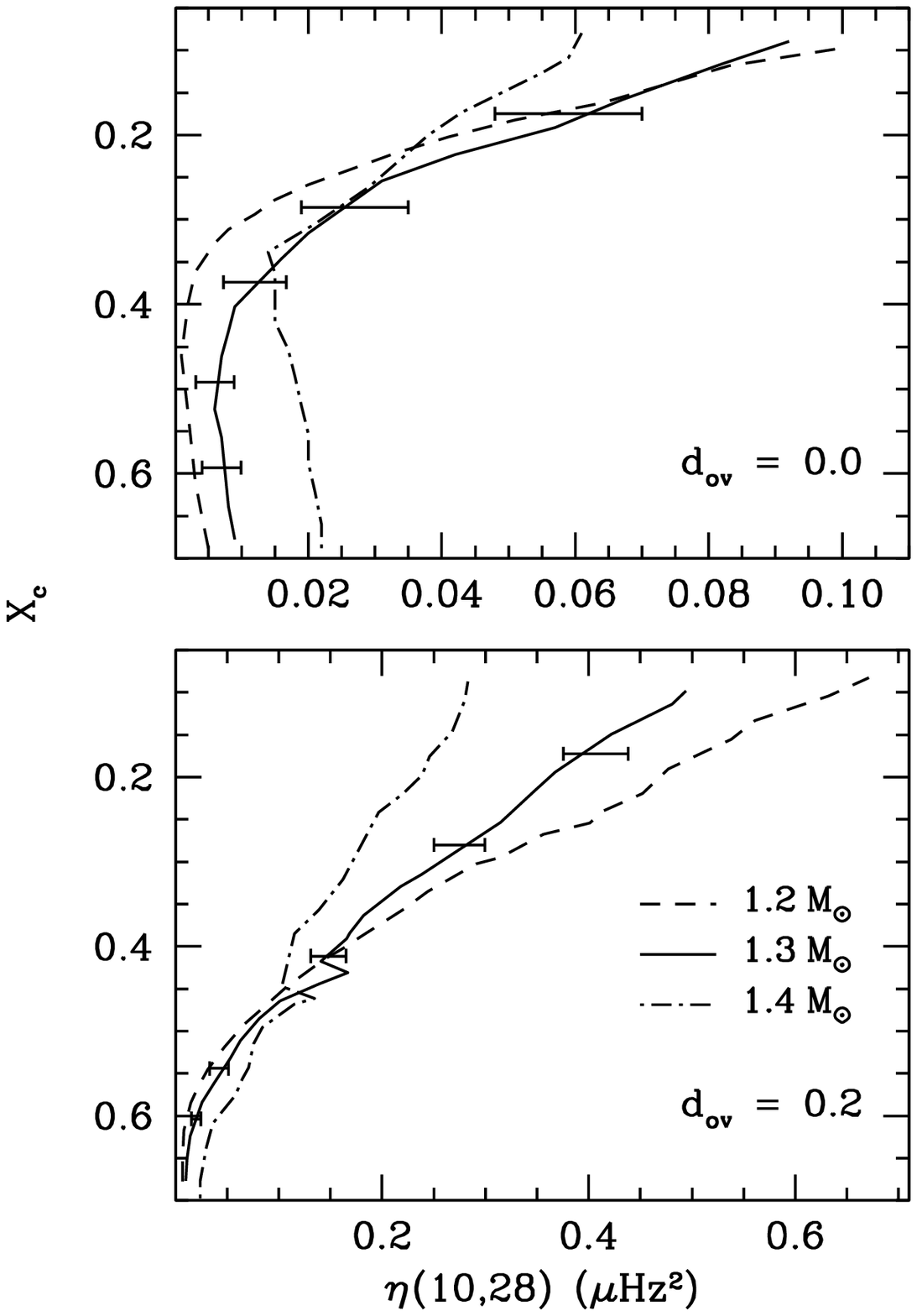}
{
The central hydrogen abundance, \xc\ is plotted as a function of the
diagnostic, \ETA, for models without overshoot ({\it upper panel}) and
with overshoot ({\it lower panel}). The errorbars represent $1\sigma$
errors in \ETA\ for relative frequency errors of $10^{-4}$, obtained
through simulations described in Section~\ref{sec:res_err_random}.  The
significant kink in the curves, especially for overshoot models,
correspond to the age of the star when the convective core stops
expanding and begins shrinking. For the no-overshoot models \ETA\ is
poorly sensitive to \xc\ at early stages of evolution and hence it will
not be possible to determine \xc\ for low values of \ETA.
}
{fig:eta_xc}
{}

\myfigure{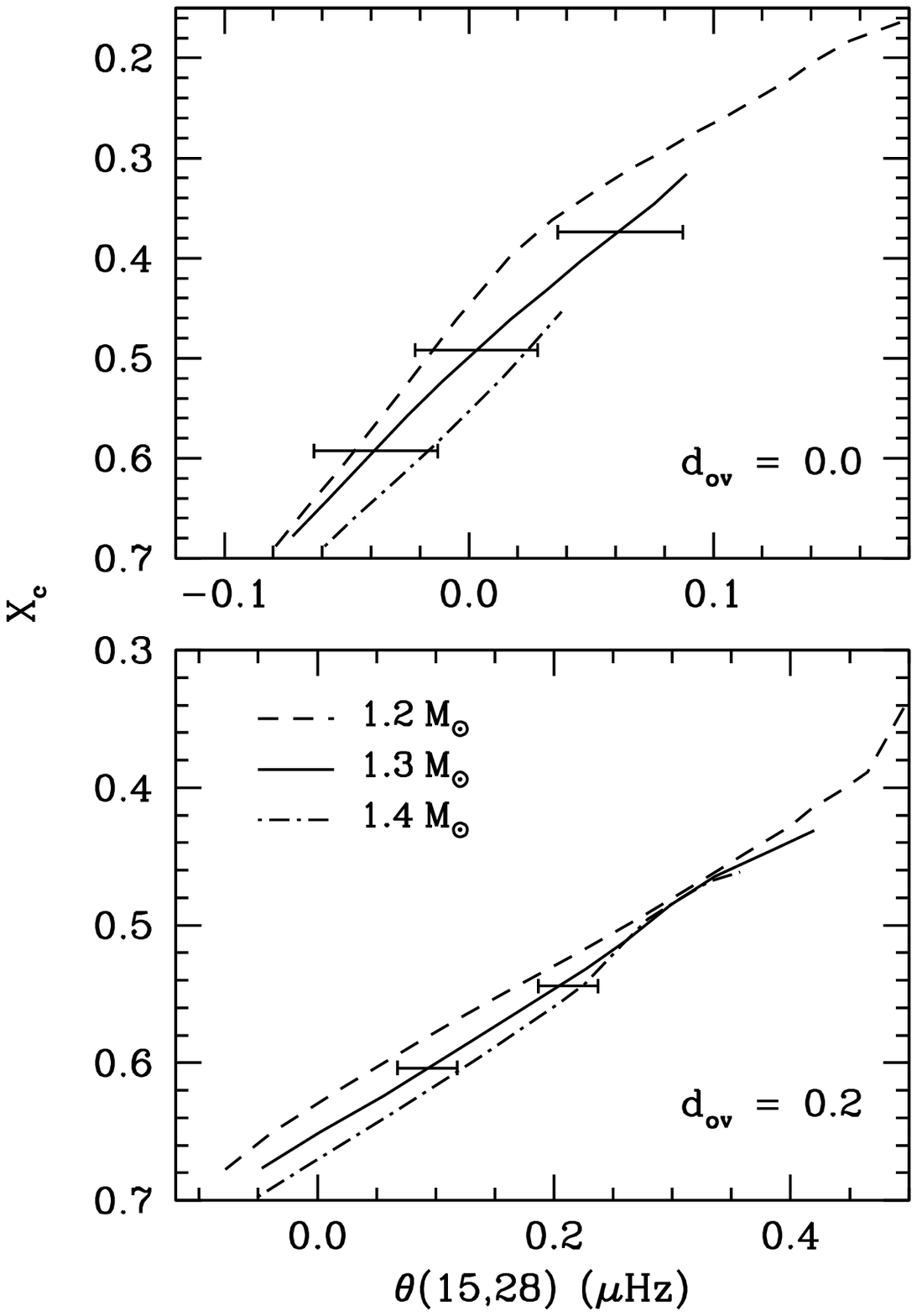}
{
The central hydrogen abundance, \xc\ is plotted as a function of the
diagnostic, \THETA, for models without overshoot ({\it upper panel}) and
with overshoot ({\it lower panel}). The errorbars represent $1\sigma$
errors in \THETA\ for relative frequency errors of $10^{-4}$, obtained
through simulations described in Section~\ref{sec:res_err_random}.  It
turns out that the correlation of \THETA\ with \xc\ becomes poor beyond
a certain evolutionary stage (depending on the mass) and the curves are
truncated only to show the part that would be usable for the
determination of \xc.
}
{fig:theta_xc}
{}

We can also use the quantity \ETA\ as an additional diagnostic for the
convective core. The variation of \ETA\ with \scmc\ is illustrated in
Fig.~\ref{fig:eta_scmc}.  We find that \ETA\ is more sensitive to the
core mass in the presence of overshoot. Actually, the use of \ETA\ as a
diagnostic is restricted only to more evolved models in the absence of
overshoot, especially for the $1.4\msun$ models. The presence of the
kink again implies that for stars near that particular evolutionary
stage the error in the core mass estimate will be larger. We also note
that unlike \DZT, the range of values of \ETA\ for a star without a
convective core, the $1\msun$ model, is restricted to much smaller
values than that for stars with a convective core. This is one advantage
that \ETA\ has, as a diagnostic of \mc, over \DZT\ (cf.
Fig.~\ref{fig:d02_scmc}). 

\subsection{Estimating \xc}
\label{sec:res_xc}

The central parts of a star change more rapidly than the outer layers
with age. Thus, it is not surprising that the small separations, or
combinations thereof, can be used as indicators of stellar age
\citep{jcd88}. We have investigated how sensitive the quantities \ETA\
and \THETA\ are to the stellar age, as measured in terms of the central
hydrogen abundance, \xc.  Figs.~\ref{fig:eta_xc} and \ref{fig:theta_xc}
show the calibration curves. As with the core mass, we find that in the
absence of overshoot \ETA\ is not very sensitive to \xc\ either.  For
overshoot models, however, \ETA\ may be used as an indicator of \xc\
throughout the main sequence phase, provided the mass is independently
known. The quantity, \THETA, on the other hand, serves as a good
diagnostic for \xc\ in the early stages of evolution. Beyond a certain
age, which depends slightly on the mass, \THETA\ no longer follows a
regular pattern with \xc. At younger ages, however, \THETA\ is less
sensitive to the stellar mass, which would make it an excellent 
indicator of the evolutionary stage 
if the mass is not well-constrained. The quantities \THETA\
and \ETA\ also serve as complementary diagnostics of \xc\ at different
stages of evolution. 

\subsection{Errors in estimation of \mc\ and \xc}
\label{sec:res_err}
\subsubsection{Random errors}
\label{sec:res_err_random}

\myfigure{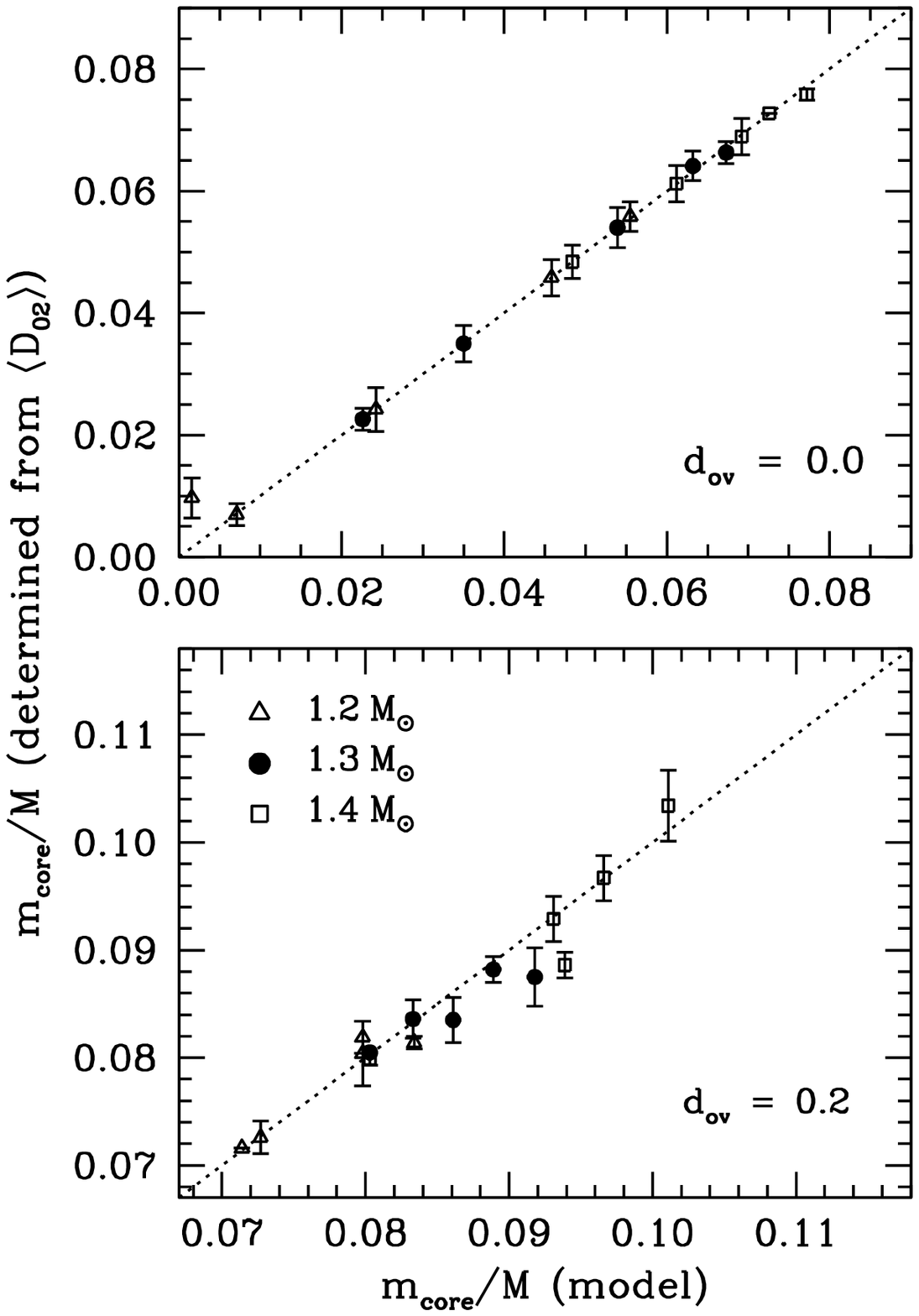}
{
The mass of the convective core, as determined through the calibration
of \scmc\ vs \DZT\ (Fig.~\ref{fig:d02_scmc}), is plotted against the
model value. Results are shown for test models both without ({\it upper
panel}) and with ({\it lower panel}) overshoot. In each case, the
calibration curve corresponding to the same mass and overshoot was used.
The dotted lines show the values which would be obtained for a perfect
match with the model.  The errorbars show the ${\mathbf 3}\sigma$ error
estimates for relative frequency errors of $10^{-4}$.
}
{fig:mc_d02res}
{}

\myfigure{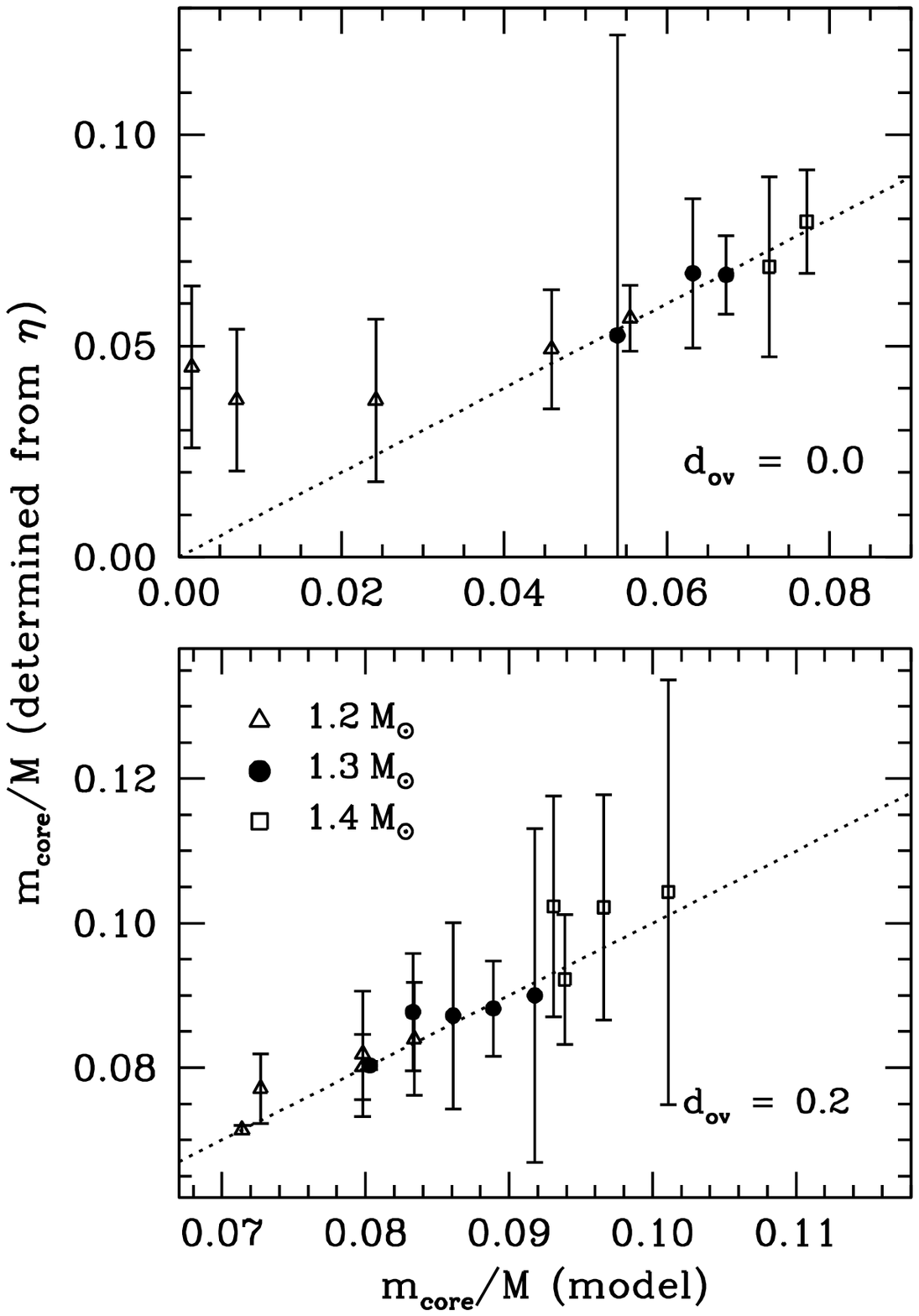}
{
The mass of the convective core, as determined through the calibration
of \scmc\ vs \ETA\ (Fig.~\ref{fig:eta_scmc}), is plotted against the
model value. Results are shown for test models both without ({\it upper
panel}) and with ({\it lower panel}) overshoot. In each case, the
calibration curve corresponding to the same mass and overshoot was used.
The dotted lines show the values which would be obtained for a perfect
match with the model.  The errorbars show the ${\mathbf 3}\sigma$ error
estimates for relative frequency errors of $10^{-4}$.
}
{fig:mc_etares}
{}

\begin{table*}
\caption{
Results of Monte Carlo simulations of determination of \mc\ and \xc\ for
5 test models at different ages with mass $M=1.3\msun$ and overshoot
$\dov = 0.2$ are shown. For each model the top row shows the original
input model values. The three subsequent rows show the results for
different quantities when relative errors in frequency of $1\times
10^{-4}$, $5\times 10^{-4}$ and $1\times 10^{-3}$ respectively are added
to the model frequencies. In each case, the results represent the
average and $1\sigma$ errors for 100 independent realisations of the
simulated data. The blank entries indicate that the technique fails to
extract the desired quantity either due to lack of sensitivity of the
diagnostic, or too high an error in frequencies to use the calibration
curves.
\label{tab:err}
}
{\scriptsize
\begin{tabular}{cr@{$\pm$}lr@{$\pm$}lr@{$\pm$}lr@{$\pm$}lr@{$\pm$}lr@{$\pm$}lr@{$\pm$}lr@{$\pm$}l}
\hline
\hline
  Age  &\multicolumn{2}{c}{$\Delta$} & \multicolumn{2}{c}{\DZT} & \multicolumn{2}{c}{\THETA} & \multicolumn{2}{c}{\ETA} & \multicolumn{4}{c}{$\mc/M$} & \multicolumn{4}{c}{\xc}        \\               
       &\multicolumn{2}{c}{        } & \multicolumn{2}{c}{    } & \multicolumn{2}{c}{      } & \multicolumn{2}{c}{    } & \multicolumn{2}{c}{using} & \multicolumn{2}{c}{using} & \multicolumn{2}{c}{using}  & \multicolumn{2}{c}{using}      \\               
 (Gyr) &\multicolumn{2}{c}{($\mu$Hz)}  & \multicolumn{2}{c}{($\mu$Hz)} & \multicolumn{2}{c}{($\mu$Hz)} & \multicolumn{2}{c}{($\mu$Hz$^2$)} &\multicolumn{2}{c}{\DZT}    & \multicolumn{2}{c}{\ETA} &\multicolumn{2}{c}{\ETA}    & \multicolumn{2}{c}{\THETA}       \\
\hline
 1.300 &  \multicolumn{2}{c}{108.862}&   \multicolumn{2}{c}{1.689}&   \multicolumn{2}{c}{0.093}&   \multicolumn{2}{c}{0.020}&   \multicolumn{4}{c}{0.083}&    \multicolumn{4}{c}{0.604}\\
 \\
       &  108.863 &  0.017 &   1.689 &  0.014 &   0.094 &  0.025 &   0.026 &  0.005 &  0.084 &  0.001 &   0.088 &  0.003 &    0.585 &  0.012 &  0.602 &  0.014 \\
       &  108.869 &  0.086 &   1.688 &  0.067 &   0.099 &  0.126 &   0.171 &  0.052 &  0.084 &  0.003 &   0.143 &  0.018 &    0.394 &  0.063 &  0.599 &  0.070 \\
       &  108.876 &  0.173 &   1.687 &  0.134 &   0.106 &  0.252 &   0.620 &  0.195 &  0.084 &  0.007 &\multicolumn{2}{c}{--  --}& \multicolumn{2}{c}{--  --}&  0.578 &  0.096 \\
\hline
 1.885 &  \multicolumn{2}{c}{102.054}&   \multicolumn{2}{c}{1.477}&   \multicolumn{2}{c}{0.212}&   \multicolumn{2}{c}{0.042}&   \multicolumn{4}{c}{0.086}&    \multicolumn{4}{c}{0.544}\\
 \\
       &  102.055 &  0.016 &   1.477 &  0.012 &   0.213 &  0.025 &   0.048 &  0.009 &  0.084 &  0.001 &   0.087 &  0.004 &    0.540 &  0.018 &  0.536 &  0.014 \\
       &  102.060 &  0.081 &   1.475 &  0.062 &   0.216 &  0.126 &   0.195 &  0.070 &  0.084 &  0.004 &   0.131 &  0.017 &    0.370 &  0.074 &  0.539 &  0.066 \\
       &  102.067 &  0.162 &   1.474 &  0.124 &   0.221 &  0.251 &   0.650 &  0.234 &  0.085 &  0.007 &\multicolumn{2}{c}{--  --}& \multicolumn{2}{c}{--  --}&  0.576 &  0.092 \\
\hline
 3.000 &  \multicolumn{2}{c}{88.636}&   \multicolumn{2}{c}{0.957}&   \multicolumn{2}{c}{0.408}&   \multicolumn{2}{c}{0.148}&   \multicolumn{4}{c}{0.092}&    \multicolumn{4}{c}{0.411}\\
 \\
       &   88.637 &  0.014 &   0.957 &  0.011 &   0.409 &  0.025 &   0.156 &  0.017 &  0.088 &  0.001 &   0.090 &  0.008 &    0.420 &  0.037 &  0.435 &  0.010 \\
       &   88.642 &  0.070 &   0.956 &  0.055 &   0.414 &  0.126 &   0.304 &  0.098 &  0.090 &  0.005 &   0.112 &  0.011 &    0.281 &  0.080 &\multicolumn{2}{c}{--  --}\\
       &   88.648 &  0.141 &   0.955 &  0.109 &   0.421 &  0.252 &   0.759 &  0.258 &  0.091 &  0.009 &\multicolumn{2}{c}{--  --}& \multicolumn{2}{c}{--  --}&\multicolumn{2}{c}{--  --}\\
\hline
 3.855 &  \multicolumn{2}{c}{78.129}&   \multicolumn{2}{c}{0.636}&   \multicolumn{2}{c}{0.505}&   \multicolumn{2}{c}{0.275}&   \multicolumn{4}{c}{0.089}&    \multicolumn{4}{c}{0.280}\\
 \\
       &   78.130 &  0.013 &   0.636 &  0.010 &   0.506 &  0.026 &   0.282 &  0.025 &  0.088 &  0.001 &   0.088 &  0.002 &    0.286 &  0.022 &\multicolumn{2}{c}{--  --}\\
       &   78.134 &  0.062 &   0.635 &  0.048 &   0.510 &  0.128 &   0.433 &  0.133 &  0.088 &  0.002 &   0.092 &  0.006 &    0.223 &  0.068 &\multicolumn{2}{c}{--  --}\\
       &   78.139 &  0.124 &   0.635 &  0.097 &   0.515 &  0.255 &   0.897 &  0.322 &  0.089 &  0.004 &\multicolumn{2}{c}{--  --}& \multicolumn{2}{c}{--  --}&\multicolumn{2}{c}{--  --}\\
\hline
 4.400 &  \multicolumn{2}{c}{71.480}&   \multicolumn{2}{c}{0.428}&   \multicolumn{2}{c}{0.550}&   \multicolumn{2}{c}{0.407}&   \multicolumn{4}{c}{0.080}&    \multicolumn{4}{c}{0.173}\\
 \\
       &   71.481 &  0.011 &   0.428 &  0.009 &   0.552 &  0.026 &   0.415 &  0.032 &  0.081 &  0.001 &   0.080 &  0.001 &    0.163 &  0.018 &\multicolumn{2}{c}{--  --}\\
       &   71.485 &  0.057 &   0.427 &  0.046 &   0.557 &  0.130 &   0.569 &  0.177 &  0.080 &  0.001 &\multicolumn{2}{c}{--  --}& \multicolumn{2}{c}{--  --}&\multicolumn{2}{c}{--  --}\\
       &   71.490 &  0.114 &   0.426 &  0.093 &   0.562 &  0.260 &   1.040 &  0.431 &  0.080 &  0.001 &\multicolumn{2}{c}{--  --}& \multicolumn{2}{c}{--  --}&\multicolumn{2}{c}{--  --}\\
\hline
\hline
\end{tabular}
}
\end{table*}

Having constructed calibration curves for the core mass and stellar age,
the next step is to check whether we can use them for real data with
errors. Since at present we do not have enough observational data to
test these methods, we undertook a Monte-Carlo simulation where random
errors with a Gaussian distribution that has a specified standard
deviation were added to the exact frequencies of a given test model. All
the necessary separations were then calculated with these error-added
frequencies and the desired quantity (\mc\ or \xc) was estimated using
the calibration curves.  Finally, the averages of the estimates along
with errorbars were obtained from 100 such realisations. Typically we
used five test models at different evolutionary stages for each mass to
check the viability of the method.  For each test model, it is assumed
that the mass of the target star is known and that the question of
overshoot has been settled independently -- through a test like
Fig.~\ref{fig:teff_d02_eta}, or by some other constraint -- so that we
can use a calibration curve constructed for the same mass and overshoot.
The systematic errors arising due to uncertainties in these respects
will be treated separately in the next section.

Figures~\ref{fig:mc_d02res} to \ref{fig:xc_thetares} show the results of
these tests. In each figure we compare the original input model value of
\mc\ and \xc\ against their estimated values. The errorbars indicate
$3\sigma$ errors obtained assuming a constant relative error of 1 part
in $10^{4}$ in frequencies. Such errors in frequencies are in line with
the expectations from the CoRoT mission. Even recent ground-based
observations have reached almost this level of precision \citep{bc02}.
For the estimation of \mc\ we note that the error is introduced at two
points -- first when we determine the scaled core mass, \scmc, from the
calibration curves (Figs.~\ref{fig:d02_scmc} and \ref{fig:eta_scmc}) and
then through the average large separation, $\Delta$ to extract \mc.  The
error in $\Delta$ is, however, tiny compared to the contribution from
the calibration curves while determining \scmc.

For the set of models with $M=1.3\msun, \dov=0.2$ we have also tested
the method for higher error in the frequencies -- relative errors of
$5\times 10^{-4}$ and $1\times 10^{-3}$ were added to the frequencies.
The results, presented in Table~\ref{tab:err}, show that the errors are
propagated linearly and that the techniques to extract \mc\ and \xc\
begin to fail only at the level of $\delta\nu/\nu \simeq 10^{-3}$, which
is ten times worse than the precision level expected from CoRoT.
However, in some cases where the diagnostic is poorly sensitive to \mc\
or \xc\ in a particular phase of evolution (e.g., \ETA\ is a poor
indicator of \xc\ at younger ages), even a small error in the diagnostic
would possibly lead to a wrong value of \mc\ or \xc\ for a star in that
phase of evolution, even if the errorbar remains small.  This is what is
seen in Table~\ref{tab:err}, for example, in the \xc\ values extracted
from \ETA\ for relative frequency errors of $5\times 10^{-4}$.

\myfigure{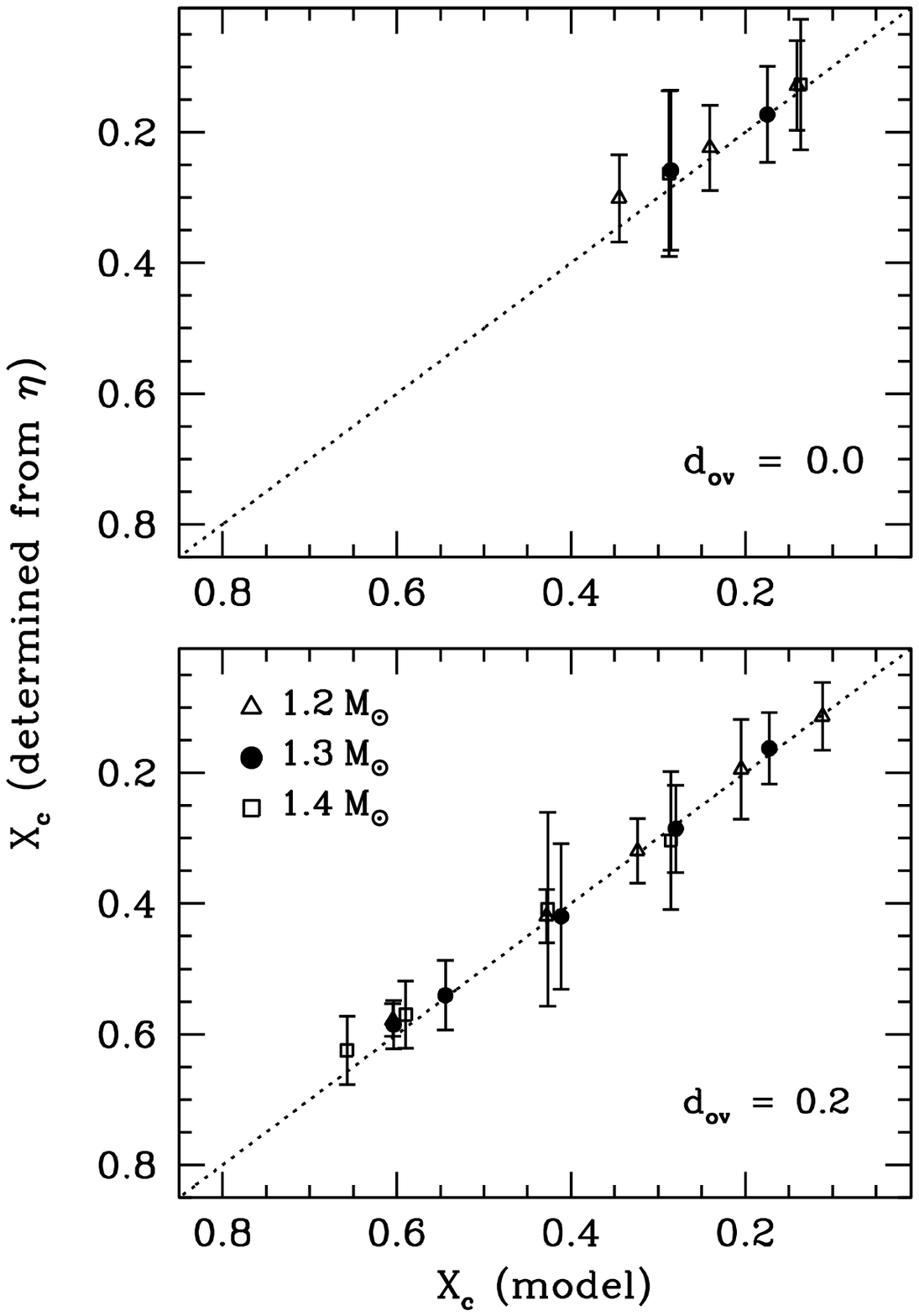}
{
The central hydrogen abundance, as determined through the calibration of
\xc\ vs \ETA\ (Fig.~\ref{fig:eta_xc}), is plotted against the model
value. Results are shown for test models both without ({\it upper
panel}) and with ({\it lower panel}) overshoot. In each case, the
calibration curve corresponding to the same mass and overshoot was used.
The dotted lines show the values which would be obtained for a perfect
match with the model.  The errorbars show the ${\mathbf 3}\sigma$ error
estimates for relative frequency errors of $10^{-4}$. The technique does
not work at younger ages for no-overshoot models due to the lack of
sensitivity of \ETA\ to \xc\ (cf.\ Fig.~\ref{fig:eta_xc}). The instances
of larger errorbars for $1.3$ and $1.4\msun$ occur when the test model
is close to the kink in the calibration curve.
}
{fig:xc_etares}
{}

\myfigure{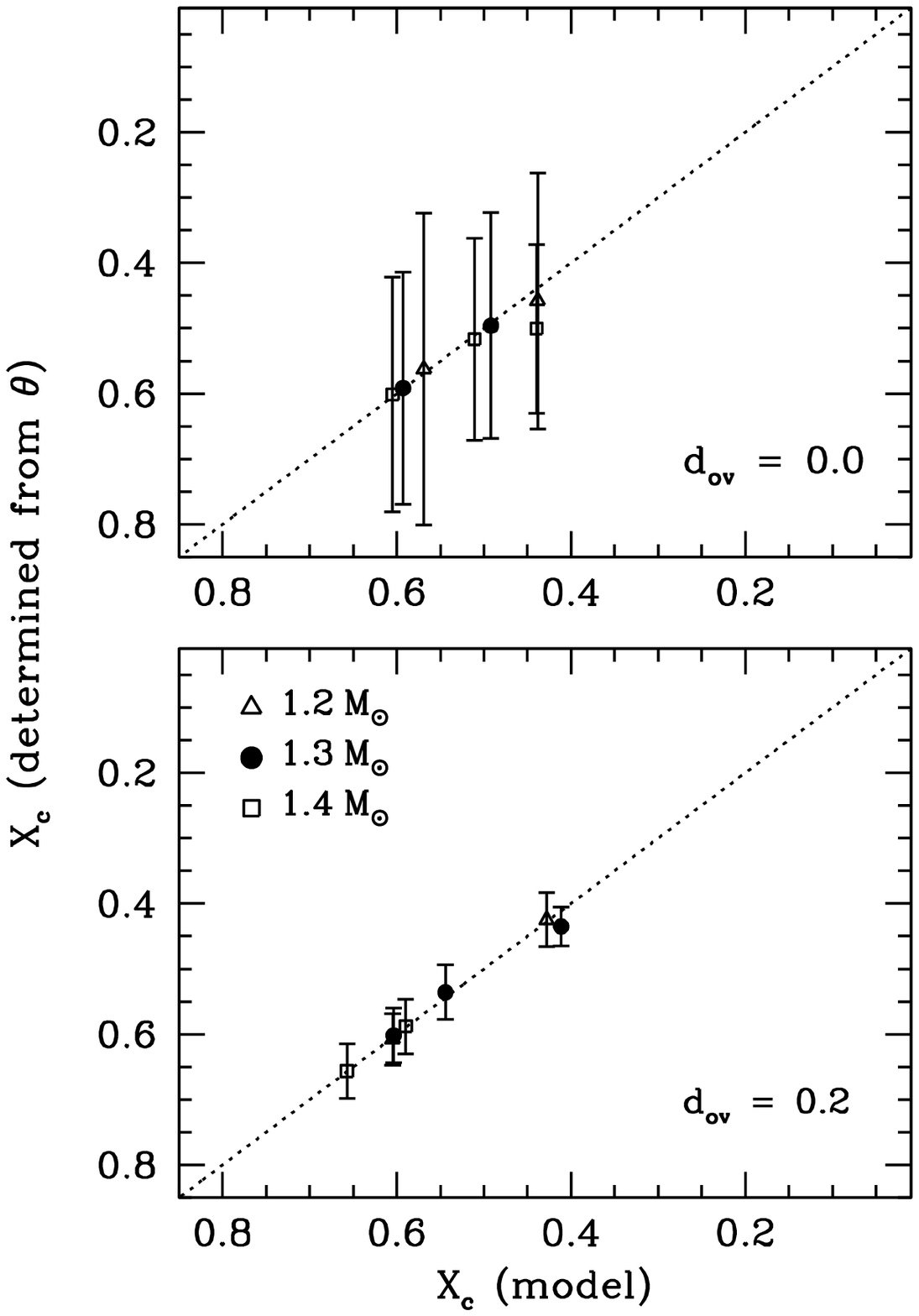}
{
The central hydrogen abundance, as determined through the calibration of
\xc\ vs \THETA\ (Fig.~\ref{fig:theta_xc}), is plotted against the model
value. Results are shown for test models both without ({\it upper
panel}) and with ({\it lower panel}) overshoot. In each case, the
calibration curve corresponding to the same mass and overshoot was used.
The dotted lines show the values which would be obtained for a perfect
match with the model.  The errorbars show the ${\mathbf 3}\sigma$ error
estimates for relative frequency errors of $10^{-4}$. The technique does
not work at low values of \xc\ due to the non-regular variation of
\THETA\ with \xc\ at more evolved stages (cf.\ Fig.~\ref{fig:theta_xc}). 
}
{fig:xc_thetares}
{}

\subsubsection{Systematic errors}
\label{sec:res_err_sys}

\myfigure{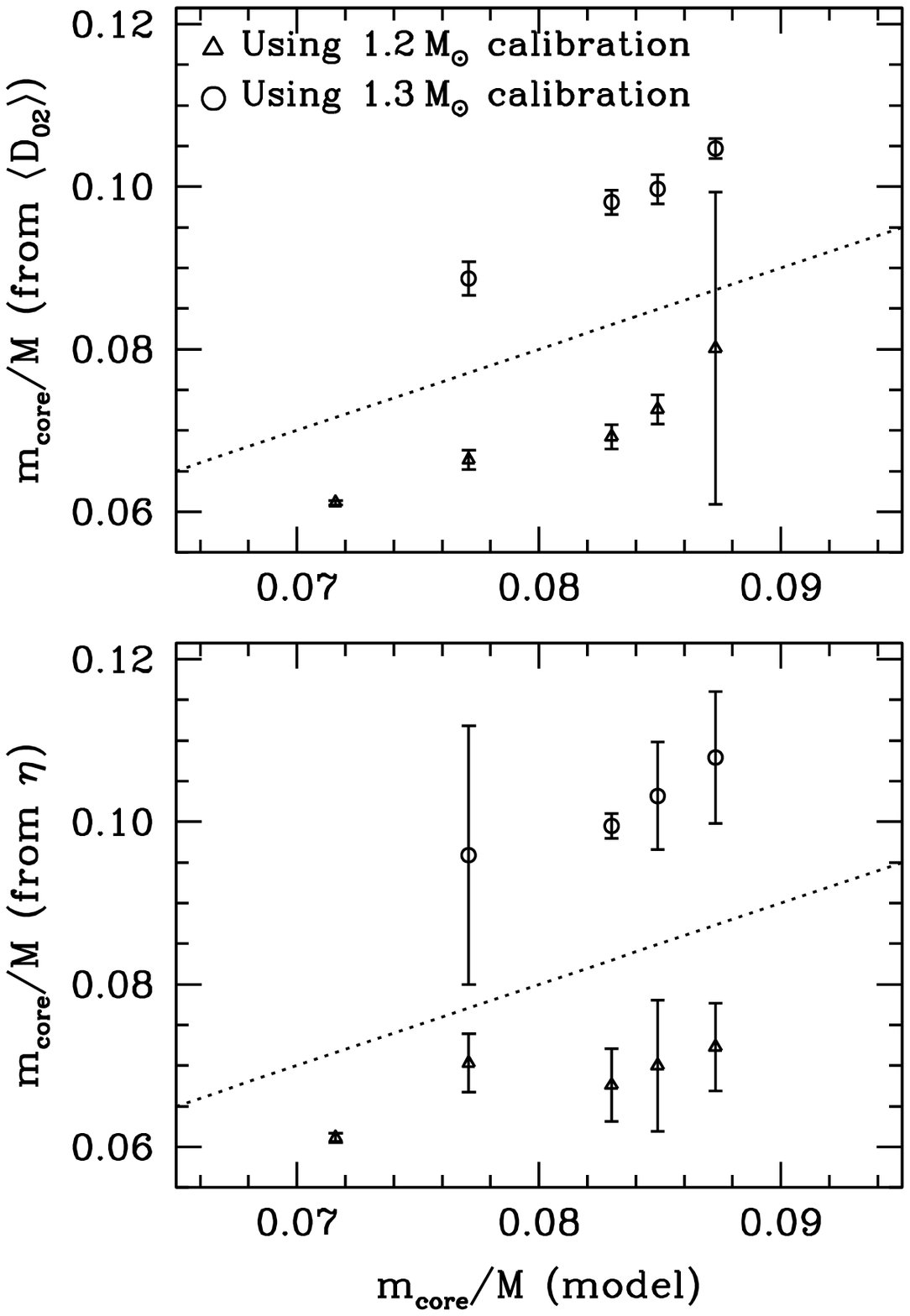}
{
The mass of the convective core of $1.25\msun$ models, as determined by
using the calibration curves of $1.2\msun$ and $1.3\msun$, is plotted
against the model value. The {\it upper panel} shows results obtained
using \DZT, while the {\it lower panel} shows results obtained using
\ETA\ calibration. The dotted lines indicate the perfect match of the
model and extracted values.  The errorbars show the random ${\mathbf
3}\sigma$ errors for relative frequency errors of $10^{-4}$. The
unusually large errorbars correspond to the uncertainty introduced by
the kink region in the calibration curves (Figs.~\ref{fig:d02_scmc}
and~\ref{fig:eta_scmc}).
}
{fig:mc_0125res}
{}

\myfigure{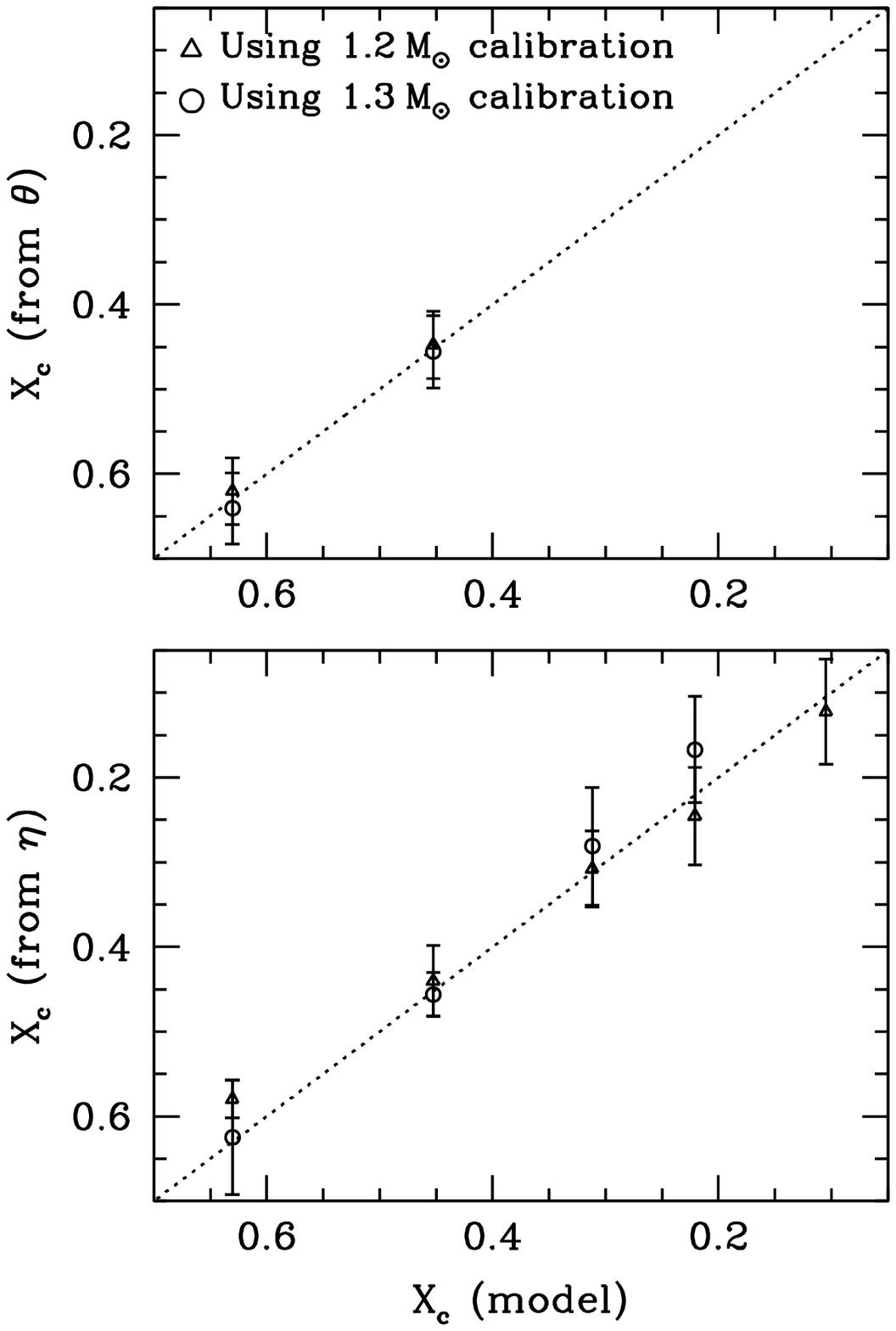}
{
The central hydrogen abundance of $1.25\msun$ models, as determined by
using the calibration curves of $1.2\msun$ and $1.3\msun$, is plotted
against the model value. The {\it upper panel} shows results obtained
using \THETA, while the {\it lower panel} shows results obtained using
\ETA\ calibration. The dotted lines indicate the perfect match of the
model and extracted values.  The errorbars show the random ${\mathbf
3}\sigma$ errors for relative frequency errors of $10^{-4}$.
}
{fig:xc_0125res}
{}

In applying our technique to determine \mc\ and \xc\ of a given target
star, we shall need to use calibration curves with approximately the
same mass and overshoot as the target. In reality, however, these
quantities are never {\it a priori} known exactly. Therefore, we need to
investigate the extent of systematic errors in \mc\ and \xc\ arising due
to the use of inexact calibration curves. 

The question of overshoot is one that asteroseismology hopes to answer.
\citet{straka05} have suggested a method of estimating overshoot through
the small separations of $g$-modes. In Section~\ref{sec:res_ovshts} we
have also demonstrated how \ETA\ can be used to detect the presence of
core overshoot. For this study, however, we realise that the calibration
curves are significantly dependent on overshoot and the systematic
errors could indeed be large if an inexact value of core overshoot is
used in the calibration ($\sim 40\%$ for an error of $\delta\dov \simeq
0.2$).  Therefore, we conclude that an independent estimate of the core
overshoot would be necessary to use our technique to determine \mc\ and
\xc\ to better accuracy.

For solar-type stars the location on the \hrd\ can be determined with a
fair accuracy which often provides a reasonable estimate of the mass of
the star. Usually, the mass of a solar-type star is known to within
$0.10\msun$ purely from its spectral type and luminosity. We have
investigated the systematic errors in \mc\ and \xc\ due to uncertainty
in the mass by using calibration curves of both $1.2\msun$ and
$1.3\msun$ for target models of $1.25\msun$. The results are shown in
Figs.~\ref{fig:mc_0125res} and~\ref{fig:xc_0125res}. The errorbars in
these figures still indicate the random errors due to relative frequency
errors of $10^{-4}$. As expected, \mc\ is systematically underestimated
when a lower mass is used for calibration, and vice versa. For an error
of $0.05\msun$ in mass, the systematic error in $\mc/M$ is $\sim 0.015$
on average. Evidently, the systematic errors in \mc\ are much larger
than the random errors arising due to errors in frequencies.

The situation is, however, different for the determination of \xc.
Fig.~\ref{fig:xc_0125res} shows that \xc\ can be determined fairly
accurately from either \THETA\ (only at younger ages) or \ETA\ even if
the mass is not known exactly. Importantly, there is no systematic trend
in the error in \xc\ due to an incorrect estimate of the mass. The
random errors actually span the range of uncertainty introduced due to
an incorrect estimate of the mass. This is not surprising, given the
closeness of the calibration curves for different masses in
Figs.~\ref{fig:eta_xc} and~\ref{fig:theta_xc}.

\section[]{Discussion}
\label{sec:discussion}

We have presented the results of our study to relate the small frequency
separations to the central layers of solar-type stars.  We have shown
that the average small separation, \DZT, can be used as a measure of the
mass of the convective core in stars of mass $1.2$--$1.4\msun$. In
addition, we also propose two combinations, \THETA\ and \ETA, of the
small separations of different pairs of degrees as new diagnostics of
the mass of the convective core, \mc\ and the stellar age, in terms of
\xc. We have tested the applicability of our technique to data with
errors by estimating the random and systematic errors in \mc\ and \xc\
through Monte Carlo simulations.  The motivation behind the use of such
combinations of \Dzt\ and \Dot\ stems from the fact that while
they both encode slightly different internal phase shifts \citep{rv00},
their combination might be sensitive to the sharp discontinuity at the
boundary of the core. A more thorough theoretical analysis of these
diagnostic quantities is required to understand their features better.
It turns out that the diagnostic value of \ETA\ and \THETA\ can be
maximised with an appropriate choice of the frequency ranges indicated
here. A different choice of the frequency range is not only feasible,
but might, in fact, be necessary depending on the availability of data.
A detailed error analysis using calibration curves with the available
observed frequencies to be used for the averaging would, however, be
necessary in order to apply the technique to real data.

It turns out that all the calibration curves depend on convective
overshoot in the core, and therefore we would need an independent
estimate of the extent of overshoot in order to determine \mc\ and \xc\
to better than $40\%$.  We find that the new diagnostic \ETA\ itself is,
however, a good indicator of the presence of overshoot.  The values of
\ETA\ are much larger in the presence of overshoot, irrespective of the
mass. 

The mass of the convective core changes as the star evolves.  For
solar-type stars with mass less than $2\msun$, it increases with age
during the first part of evolution on the main sequence before shrinking
down slowly in the later part. The values of the small separations,
however, decrease monotonically with age, being more sensitive to the
general evolution than to the size of the core. We attempt to account
for the general change in the mean density with age by scaling the mass
of the convective core with the large separation. We find that the
average small separation, or the new diagnostic \ETA, varies almost
monotonically with the scaled core mass, \scmc.  This scaling enables us
to calibrate \DZT\ or \ETA\ against the core mass. It turns out that the
random error in determining \mc\ through \DZT\ is smaller than that
through \ETA. On the other hand, the systematic errors due to inaccuracy
in the total mass of the target star in the calibration curves can be as
large as up to $20\%$ for typical mass errors of $10\%$.

We have also shown that the new diagnostic quantities, \ETA\ and \THETA\
may be used to estimate the age of a main sequence star. We find that
they serve as nearly complementary indicators of the central hydrogen
abundance -- while \THETA\ is a fairly good indicator of \xc\ during the
early phase of evolution, \ETA\ can be used at more evolved stages. Both
of these quantities, as diagnostics of \xc, are fairly insensitive to
mass, which make them better indicators of the evolutionary stage of a
star than the small separations themselves. The actual age of the star
in terms of years would, of course, still depend heavily on the stellar
mass.  The random errors due to errors in frequency actually dominate
the systematic errors due to uncertainty in mass. \xc\ can be determined
to $5\%$ for typical frequency errors of $\sim 0.1\mu$Hz. We note that
\ETA\ is more sensitive to both \mc\ and \xc\ when overshoot is included
in the models. 

While the age of the star determined through \ETA\ or \THETA\ is quite
independent of stellar mass, the determination of the mass of the
convective core depends on the total mass of the star. Further, both of
these techniques are sensitive to the extent of convective overshoot in
the core. We have, in this work, provided a method to test the extent of
overshoot through the diagnostic \ETA\ itself. The estimate of the
stellar mass, however, needs to be obtained independently. For the
fortunate cases of binary stars \citep[e.g., as in
$\alpha$~Cen,][]{pnn99}, the mass is usually known to a high accuracy
from dynamic considerations. Even for single stars, a precise location
on the \hrd\ and a good estimate of the metallicity often helps to
constrain the mass. In any case, a detailed correlation analysis between
the age, the stellar mass and the extent of overshoot will be needed to
correctly estimate the uncertainty in the results obtained from this
technique. 

\section*{Acknowledgements}

BLC was partially supported by NSF grant ATM-0348837 to SB.
PD was supported by NASA grant NAG5-13299.

\bsp

\label{lastpage}

\end{document}